\documentclass[]{iise}

\usepackage{multirow}
\usepackage{graphicx}
\usepackage{wrapfig}
\usepackage{placeins}

\conference{Proceedings of the IISE Annual Conference \& Expo 2026\\
Y. Xiang, D., Yu, R. Thiesing, eds.}

\title{\titlesize Optimal Scheduling of Road Maintenance Jobs Considering Impact on Traffic Flows}

\author{
No Author 1 Yet\\No Organization Yet\\No Location Yet \\
\vspace{0.3cm}
No Author 2 Yet\\No Organization Yet\\No Location Yet}
\authorlist{Nandepu, Kalepu, Ciavarella and Park}
\abstractID{1234}

\begin{document}

\author{Charitha Nandepu, Lohitha Kalepu, Gabriele Ciavarella and SangWoo Park}

\maketitle

\begin{abstract}
\vspace{-5mm}
Network-level maintenance planning requires repeated evaluations of equilibrium traffic flows under road capacity reductions. While equilibrium traffic assignment models are well established, their repeated solution quickly becomes computationally prohibitive and challenging to embed within maintenance scheduling problems. This paper investigates data-driven surrogate models that approximate equilibrium arc flows directly from origin–destination demand, using optimization-based equilibrium solutions as ground truth. A real-world case study based on traffic data from the Newark, New Jersey area demonstrates the effectiveness of the proposed approach as a scalable building block for future maintenance scheduling frameworks.
\end{abstract}
\vspace{-6mm}
\section*{Keywords}
Optimal scheduling, road maintenance, traffic data, network equilibrium flow, deep learning
\vspace{-4.0mm}
\section{Introduction and Background} \label{sect:introduction}

Road maintenance is a critical component of infrastructure management, yet poorly planned interventions often trigger network-wide traffic disruptions and economic losses. Even localized lane closures can generate non-local effects as drivers reroute via navigation tools like Google Maps or Apple Maps, which amplify the systemic impact of congestion. Traditional maintenance scheduling models often embed traffic assignment within larger optimization frameworks, requiring equilibrium flows to be recomputed for every candidate configuration---a process that creates a significant computational bottleneck~\cite{Liu2020}.\
While classical traffic assignment is rooted in Wardrop's user equilibrium principle and the Beckmann formulation~\cite{Beckmann75,LeBlanc75}, existing studies often suffer from simplifying assumptions, such as focusing on isolated corridors or assuming exogenous, fixed traffic flows. These assumptions become increasingly unrealistic for large-scale networks where traffic patterns respond endogenously to capacity reductions. Recent transportation research has sought to address these challenges through computationally efficient user-equilibrium algorithms~\cite{Nikolic2021}.\
In real-world scenarios, infrastructure managers must schedule multiple projects over a finite horizon while forecasting these complex flow responses. This network-level maintenance scheduling problem aims to minimize total user delay costs, acknowledging that traffic patterns depend on both the timing of activities and road-specific characteristics. Recent developments in graph neural networks and surrogate modeling offer a promising alternative by learning complex network operators directly from data~\cite{Wu2020,Makarov2024}. Motivated by these advancements, our approach utilizes surrogate modeling to approximate nonlinear mappings in network flows, replacing computationally intensive optimization in infrastructure planning.\

\vspace{-4.5mm}
\section{Problem Statement}
From an operational perspective, evaluating the impact of candidate maintenance schedules therefore requires repeatedly computing equilibrium traffic flows under different combinations of maintenance activities and demand conditions. While equilibrium traffic assignment models are well established and can be formulated as convex optimization problems, their repeated solution across multiple time periods and maintenance configurations quickly becomes computationally burdensome, especially in network-level settings.\
In this context, the objective of this work is not to solve the full maintenance scheduling problem end-to-end. Instead, we focus on a key computational subproblem that arises within such frameworks: the efficient evaluation of equilibrium traffic responses under maintenance-induced capacity changes. Specifically, we study how equilibrium traffic flows can be accurately approximated using data-driven models trained on optimization-based equilibrium solutions.\
The proposed approach is developed and validated through a real-world case study based on traffic data from the road network surrounding the city of Newark, New Jersey. 

\vspace{-4.5mm}
\section{Approach}\label{sect:approach}
This section presents a comprehensive case study evaluating the proposed modeling framework on a real-world traffic network derived from StreetLight data. The objective is twofold: (i) to generate equilibrium traffic states using a convex optimization-based user equilibrium formulation, and (ii) to assess the ability of learning-based models to approximate these equilibrium flows directly from Origin--Destination (OD) demand inputs. The equilibrium solutions serve as ground truth for training and evaluating data-driven models that enable fast inference within a maintenance scheduling context.
Specifically, equilibrium traffic flows are generated for each day–hour Origin–Destination demand scenario by solving a user equilibrium traffic assignment problem based on Wardrop’s principle. The resulting equilibrium arc flows represent stable traffic patterns in which no traveler can reduce travel time by unilaterally changing routes.
\vspace{-4.5mm}
\subsection{Traffic Dataset and Preprocessing}
The data used for this study was obtained from a traffic data analytics platform, StreetLight Data. StreetLight's Network Origin-Destination Analysis tool was used to get Origin-Destination (OD) traffic volumes. These metrics are deduced from aggregated GPS-based trip samples and scaled to OD volumes using the StreetLight Network Origin--Destination methodology~\cite{NetworkODMethodology}.\
For this study, we have selected a network of 49 road segments within the metropolitan city of Newark, New Jersey, with the highest traffic volume (identified as the Garden State Parkway corridor), where maintenance activities result in higher congestion and delays. 
Treating each road segment as an origin and destination, the OD matrix was constructed for every hour of the day and week throughout the month of October 2025. 
Traffic data for certain hours of the month were unavailable and are entered with zeros in the OD matrix to maintain consistency. Additionally, since the region is only a part of the city, some roadway links are truncated at the boundary, giving rise to new pairs called “Other Origins” and “Other Destinations” by StreetLight. To address this, we aggregated all “Other Origins” and “Other Destinations” as a single external origin and destination node. 
With the inclusion of these external nodes and the OD pairs, the final OD representation becomes a square 50x50 matrix. For further computational feasibility, this OD matrix is converted into a 3D JSON tensor with dimensions 7 x 2500 x 24, representing the flows for 2500 flattened OD pairs for 24 hours a day for all 7 days of the week. 

\vspace{-4.0mm}
\subsection{Computing Network Equilibrium Flows}
\begin{wrapfigure}[15]{r}{0.45\textwidth}
\vspace{-4mm}
\centering
\includegraphics[width=0.43\textwidth]{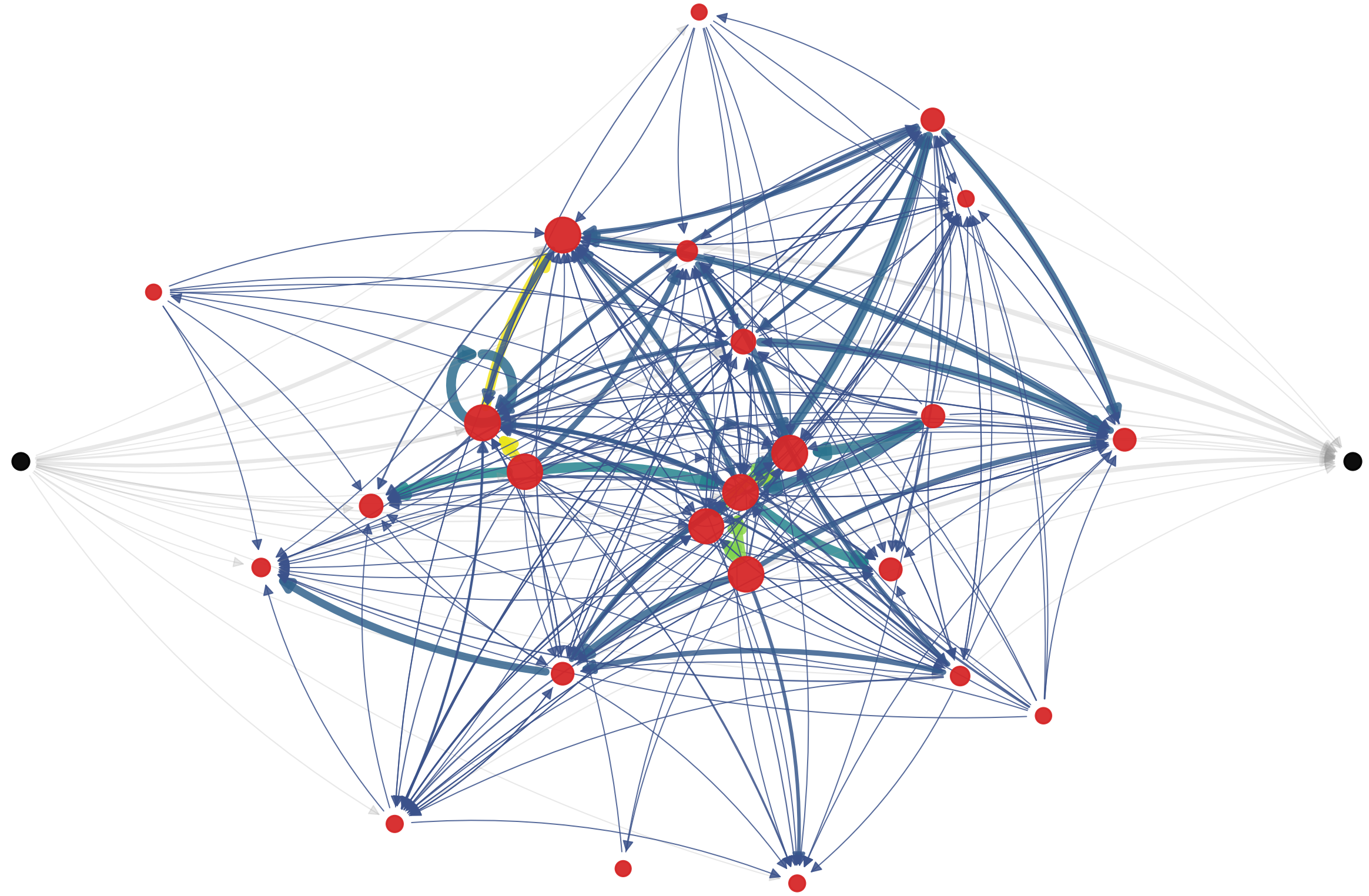}
\vspace{-3mm}
\caption{\footnotesize Equilibrium OD flow visualization for Day~1, Hour~8. Edge width reflects flow magnitude, while color encodes log-scaled flow intensity, with brighter colors indicating higher equilibrium flows.}
\vspace{-2mm}
\label{fig:eq_flow_network}
\end{wrapfigure}
The transportation network is modeled as a directed graph in which nodes represent traffic analysis zones and directed arcs represent feasible traffic links between zones. Each arc captures directional traffic movement and serves as
the fundamental unit for equilibrium flow computation. For each day--hour scenario, travel demand is represented by an Origin--Destination (OD) matrix, where each entry denotes the demand from an origin zone to a destination zone. Self-demand entries are set to zero. These OD matrices are derived from StreetLight Data and scaled into model units prior to optimization.
Arc-level parameters are obtained from an externally generated dataset containing calibrated coefficients for all valid arc pairs. For each arc $k=(i,j)$, two parameters are specified: a free-flow travel time coefficient $a_{ij}$ and a congestion sensitivity coefficient $b_{ij}$. These parameters define the nonlinear travel time function,
$t_{ij}(x_{ij}) = a_{ij} + b_{ij}x_{ij}^{4}$,
where $x_{ij}$ denotes the total flow on arc $(i,j)$.
For each day--hour scenario, equilibrium arc flows are computed using a Beckmann-type user equilibrium formulation. The optimization problem is implemented in CVXPY and solved using the MOSEK solver, which provides robust performance for large-scale convex programs. The formulation is based on a destination-based flow decomposition. Let $X_{k,s}$ denote the flow on arc $k$ associated with trips destined for node $s$. The total equilibrium flow on arc $k$ is defined as  
$x_k = \sum_s X_{k,s}.$

All decision variables are constrained to be nonnegative. The objective minimizes the total integrated travel cost:
\begin{equation}
\min_X \sum_k \left( a_k x_k + \frac{b_k}{5} x_k^5 \right),
\end{equation}
Flow conservation constraints are imposed for every destination and every node. Let $D(j,s)$ denote the OD demand from origin node $j$ to destination node $s$. For each destination $s$, flow conservation is enforced such that, at non-destination nodes, the difference between outgoing and incoming flows equals $D(j,s)$, while at destination nodes, incoming flow minus outgoing flow equals the total terminating demand associated with destination $s$.

For each day--hour scenario, equilibrium arc flows are generated through a structured data processing and optimization pipeline. The OD demand data are loaded from the weekly JSON file derived from StreetLight’s OD estimation framework, reshaped into square matrices, scaled into model units, and cleaned of self-demand. Network structure and arc parameters are loaded from the CSV file containing calibrated coefficients. For each OD matrix, equilibrium flows are computed by solving the destination-decomposed user equilibrium problem. The solution is strictly convex and uniquely determined.

Once the equilibrium solution is obtained, total arc flows are extracted and converted back into physical units (vehicles per hour). Figure~\ref{fig:eq_flow_network} visualizes equilibrium arc flows for a representative day--hour scenario. Travel times are evaluated at equilibrium using the nonlinear arc cost functions. All equilibrium arc flows are aggregated into a single dataset and used as ground truth for training and evaluating learning-based models.

\vspace{-4.0mm}
\subsection{Learning Task Definition and Data Splitting}
The learning objective is to approximate the nonlinear mapping from OD demand matrices to equilibrium arc flow vectors generated by the optimization model described in Section 3.2. In other words, we want to establish:
$f_{\theta}: \text{OD}^{(d,h)} \rightarrow \mathbf{x}^{(d,h)}.$
Here, $\text{OD}^{(d,h)}$ denotes the OD matrix for day $d$ and hour $h$, and $\mathbf{x}^{(d,h)}$ denotes the corresponding equilibrium arc flows. OD matrices are flattened into vectors, and outputs are aligned using a consistent global arc index. To rigorously evaluate generalization, the scenario corresponding to Day~1, Hour~1 is completely excluded from training, validation, and testing and is reserved exclusively for inference.

\vspace{-4.0mm}
\subsection{Neural Architectures and Experimental Design}

Four neural architectures are evaluated in this study: a Multilayer Perceptron (MLP), a Convolutional Neural Network (CNN), a Graph Neural Network (GNN), and an attention-based neural network. Each model is trained to approximate the mapping from an Origin--Destination (OD) demand matrix to equilibrium arc flows computed by a traffic assignment solver.
The MLP operates on flattened OD demand vectors and predicts equilibrium arc flows using a sequence of fully connected layers. This architecture learns a global nonlinear mapping between OD demand and arc-level flows and serves as a baseline for comparison.

\begin{wraptable}[8]{r}{0.47\textwidth}
\vspace{-12pt}
\centering
\caption{Performance comparison of models for equilibrium flow prediction}
\vspace{-2mm}
\label{tab:model_comparison}
\small
\setlength{\tabcolsep}{3pt} 
\renewcommand{\arraystretch}{0.9} 

\begin{tabular}{lccc}
\hline
\textbf{Model} 
& \textbf{RMSE} 
& \textbf{$R^2$} 
& \textbf{High-flow RMSE} \\
& \textbf{(veh/hour)} 
& 
& \textbf{($>200$ veh/hour)} \\
\hline
MLP              & 84.32  & 0.9088 & 605.93 \\
CNN              & 133.84 & 0.7703 & 983.82 \\
GNN              & 58.48  & 0.9561 & 167.20 \\
Attention-based NN & 35.87 & 0.9835 & 236.22 \\
\hline
\end{tabular}
\vspace{-10pt}
\end{wraptable}
The CNN processes the OD demand matrix as a two-dimensional input and applies convolutional layers followed by fully connected layers to generate arc flow predictions. The OD matrix is treated as a single-channel grid, and convolutional feature extraction is applied uniformly across the matrix.
The GNN represents the transportation system as a directed graph, with zones modeled as nodes and arcs modeled as edges. Node features are constructed from aggregated OD demand statistics, while edge features include OD demand and static arc parameters. Message passing is used to propagate information across the network before producing arc-level flow predictions.
The attention-based neural network represents each OD pair as a token and applies self-attention to learn interactions among all OD pairs within a scenario. The resulting global representation is mapped to equilibrium arc flows using a feedforward decoder.
All models are trained using scenario-level splits to avoid information leakage across day--hour combinations. The scenario corresponding to Day~1, Hour~1 is excluded entirely from training, validation, and testing and is reserved exclusively for out-of-sample inference. Model performance is evaluated using RMSE and $R^2$ in veh/hour, with additional evaluation focused on arcs exceeding 200~veh/hour.

\vspace{-2mm}
\section{Results}
Performance is evaluated using RMSE and $R^2$, with special attention to high-flow arcs exceeding 200~veh/hour. This threshold isolates primary loading corridors from background flow, focusing evaluation on the operationally significant arcs that most influence maintenance planning. The attention-based model achieves the strongest performance, explaining approximately 98\% of flow variance with the lowest overall RMSE. The GNN also performs strongly, particularly on high-demand arcs, while the MLP and CNN exhibit significantly higher errors on these critical links. These results demonstrate that capturing global interactions is essential for accurately approximating equilibrium patterns. Specifically, by ensuring accuracy on high-demand corridors, the model provides the reliable delay estimations necessary for effective system-wide planning.
\FloatBarrier
\section{Conclusion}
This paper presented a modeling framework for optimizing road maintenance scheduling while accounting for endogenous traffic flow responses at the network level. A convex Beckmann-type formulation was used to compute equilibrium arc flows from real-world OD demand derived from StreetLight data, which served as ground truth for training learning-based Deep Learning models. Among the evaluated models, the attention-based neural network achieved the highest predictive accuracy and most faithfully reproduced network congestion patterns.

\bibliography{references}

\end{document}